\documentclass[twocolumn,10pt,amsmath,amssymb,showkeys,aps,nofootinbib]{revtex4-2}

\usepackage{graphicx}
\usepackage{dcolumn}
\usepackage{bm}
\usepackage{hyperref}
\usepackage{subfigure}
\usepackage{cleveref}
\usepackage{nicefrac}

\usepackage{xcolor}

\usepackage{multirow} 

\begin{document}

\title{Spectral Bifurcations in Quasinormal Modes of Regular BTZ Black Holes}

\author{Kartheek Hegde}%
\email{hegde.kartheek@gmail.com}
\affiliation{Department of Physics, National Institute of Technology Karnataka, Surathkal 575 025, India}%

\author{Tajron Juri\'c}
\email{tjuric@irb.hr}
\affiliation{Rudjer Bo\v{s}kovi\'c Institute, Bijeni\v cka  c.54, HR-10002 Zagreb, Croatia}

\author{A. Naveena Kumara}
\email{naviphysics@gmail.com}
\email{nathith@irb.hr}
\affiliation{Rudjer Bo\v{s}kovi\'c Institute, Bijeni\v cka  c.54, HR-10002 Zagreb, Croatia}

\begin{abstract}

We study the quasinormal spectrum of massless scalar fields propagating on a family of regular BTZ black holes arising from an infinite tower of dimensionally regularized Lovelock corrections. These geometries are asymptotically AdS, reduce to the standard BTZ solution in the limit $\ell \to 0$, and resolve the central singularity by introducing a smooth core controlled by the new length scale $\ell$. The scalar quasinormal modes are computed using both Leaver's continued–fraction method and the Horowitz--Hubeny power–series method; the two approaches agree to high accuracy across the parameter space. We find that the regularization preserves linear stability ($\omega_I < 0$) while qualitatively reshaping the spectrum: as $\ell$ increases, BTZ–like complex branches collide with the imaginary axis and undergo a hierarchy of bifurcations into multiple purely imaginary branches, leading to mode switching and a nontrivial reordering of overtones as functions of $\ell$ and the harmonic index $m$. Our results place regular BTZ black holes within the emerging family of bifurcating quasinormal spectra known from nearly extremal and asymptotically AdS black holes, and highlight these $(2+1)$-dimensional geometries as a controlled arena for exploring geometric mechanisms behind spectral branching and late–time ringdown in regular black hole spacetimes.

\end{abstract}

\keywords{Quasinormal modes, Regular BTZ black holes, Spectral bifurcations, Black hole perturbations, Lovelock gravity}
\maketitle

\section{Introduction}\label{sec1}

The landmark detection of gravitational waves by the LIGO, VIRGO, and KAGRA collaborations has ushered in a new era of observational astronomy and fundamental physics~\cite{Berti:2025hly, LIGOScientific:2016aoc, KAGRA:2022twx}. Black hole mergers and their post-merger ringdown phases provide a unique window into the strong-field regime of gravity. These ringdown signals are dominated by quasinormal modes (QNMs), which describe the characteristic oscillations of a perturbed black hole spacetime and are determined solely by the black hole's background geometry and field content~\cite{Vishveshwara:1968ksg}. As such, QNMs play a crucial role in testing general relativity (GR), probing the no-hair conjecture, and constraining modified theories of gravity. Moreover, they serve as an essential diagnostic tool in the quest to understand how black holes respond to external perturbations, making their study highly relevant across a range of contexts, from astrophysical observations to holographic dualities and beyond~\cite{Kokkotas:1999bd, Nollert:1999ji, Ferrari:2007dd, Berti:2009kk, Konoplya:2011qq, Berti:2018vdi, Cardoso:2019rvt}.

The computation and analysis of QNMs have historically been carried out within the classical framework of GR. However, at high curvatures near black hole singularities, quantum gravity effects or higher-order corrections are expected to become significant. This has motivated the search for effective descriptions of gravity that go beyond Einstein’s theory, especially in the form of higher-derivative extensions such as Lovelock~\cite{Lovelock:1971yv}, Gauss-Bonnet~\cite{Fernandes:2022zrq}, and quasi-topological gravity~\cite{Hennigar:2017ego}. In this context, the study of QNMs offers a natural and robust arena to probe the viability and physical predictions of such corrections. In particular, QNMs can help distinguish between black holes that are merely deformations of GR solutions and those that exhibit qualitatively new features, such as regular (non-singular) cores or modified causal structures.

One of the most profound limitations of classical GR is its prediction of spacetime singularities, such as those occurring at the center of black holes or at the Big Bang~\cite{Penrose:1964wq, Hawking:1970zqf}. These singularities signal the breakdown of the classical description of gravity, leading to geodesic incompleteness and the divergence of physical quantities such as curvature invariants. Over the past few decades, a wide variety of \emph{regular black hole models} have been proposed to resolve these singularities. These include constructions using non-linear electrodynamics, loop quantum gravity, and effective energy-momentum tensors engineered to yield non-singular geometries (see \cite{Bambi:2023try} and references therein). However, many of these models suffer from theoretical drawbacks: they often violate energy conditions, rely on ad hoc matter content, or introduce fine-tuned relationships between the parameters of the solution and those of the theory. Moreover, such solutions frequently exhibit dynamical instabilities, particularly near their inner (Cauchy) horizons~\cite{Poisson:1990eh, DiFilippo:2022qkl}.

A recent and promising alternative is the approach based on an \emph{infinite tower of higher-curvature corrections}, motivated by effective field theory and quantum gravity frameworks~\cite{Bueno:2024dgm}. These corrections naturally emerge in low-energy limits of string theory~\cite{Gross:1986mw, Gross:1986iv, Grisaru:1986vi}, asymptotically safe gravity~\cite{Bonanno:2020bil, Knorr:2022dsx, Knorr:2018kog, Dupuis:2020fhh}, and other ultraviolet (UV) completions of GR. In higher dimensions, quasi-topological gravities, built from special combinations of Riemann tensor invariants, offer an appealing playground for such theories~\cite{Aguilar-Gutierrez:2023kfn, Ahmed:2017jod, Bueno:2016lrh, Bueno:2019ycr, Bueno:2022res, Bueno:2017qce, Hennigar:2016gkm, Hennigar:2017ego, Moreno:2023rfl, Moreno:2023arp}. Importantly, when evaluated on symmetric backgrounds, these theories can retain second-order field equations and avoid the Ostrogradsky instabilities that typically affect higher-derivative models.

Within this framework, it has been demonstrated that black holes can dynamically form with fully regular interiors and avoid mass inflation instabilities, provided an infinite tower of higher-order curvature corrections is taken into account~\cite{Bueno:2024zsx, Bueno:2024eig}. This leads to what has been termed geometric inflation in cosmological settings~\cite{Arciniega:2018tnn, Arciniega:2018fxj, Arciniega:2020pcy, Edelstein:2020nhg, Jaime:2021pqn, Arciniega:2025viq}. These models do not require exotic matter sources and instead rely solely on geometric corrections to the Einstein-Hilbert action. However, a key limitation is that such quasi-topological theories are well-defined only in ($ D \geq 5$ ) dimensions, and their four-dimensional or lower-dimensional counterparts typically suffer from ill-defined limits or yield trivial corrections due to the topological nature of Lovelock terms in low dimensions.

To circumvent this, dimensional regularization techniques have been proposed as a way to define well-behaved scalar-tensor theories in four dimensions by applying a limiting procedure to the Lovelock invariants~\cite{Fernandes:2025fnz}. The resulting scalar-tensor theories belong to the Horndeski class and admit well-defined variational principles and second-order field equations. In particular, based on the approach of Refs. \cite{Hennigar:2020fkv, Hennigar:2020drx}, recent work by Fernandes~\cite{ Fernandes:2025eoc} has extended this procedure to all Lovelock orders in (2+1)-dimensional spacetimes, leading to a class of scalar–tensor theories that admit everywhere regular BTZ black hole solutions. These black holes preserve many of the desirable features of the Bañados–Teitelboim–Zanelli (BTZ) solution~\cite{Banados:1992wn, Carlip:1995qv, Banados:1992gq}, such as AdS asymptotics and thermodynamic structure, while replacing the central singularity with a smooth core and introducing an extremal inner horizon that protects against mass inflation instabilities.

The motivation to study BTZ black holes in three-dimensional AdS gravity goes beyond mere tractability. Originally discovered in 1992, the BTZ black hole represents a rare example of an exact, analytically tractable black hole solution with thermodynamic and causal properties remarkably similar to those of higher-dimensional black holes. Despite the absence of local degrees of freedom in three-dimensional gravity, the BTZ black hole exhibits genuine event horizons, a well-defined Hawking temperature, and a Bekenstein–Hawking entropy. Furthermore, BTZ black holes play a central role in the AdS/CFT correspondence, where they are dual to thermal states in two-dimensional conformal field theories~\cite{Carlip:2005zn, delaFuente:2013nba}. As such, they serve as ideal laboratories for studying gravitational perturbations, holographic transport, and quantum aspects of black holes in simplified settings.

In recent years, a large number of studies have examined QNMs of scalar, vector, and spinor fields in BTZ and BTZ-like backgrounds~\cite{Cardoso:2001hn, Panotopoulos:2018can, DalBoscoFontana:2023syy, Cheriyodathillathu:2022fwe, Katagiri:2022qje, Chernicoff:2020kmf, MahdavianYekta:2018lnp, Prasia:2016esx, Datta:2011za, Kwon:2010um, Denef:2009kn, Kuwata:2008in, Gupta:2005sf, Birmingham:2003wa, Crisostomo:2004hj, Birmingham:2001hc, Birmingham:2001pj, Gupta:2015uga, Gupta:2017lwk}, often yielding analytical solutions in terms of hypergeometric or Heun functions. Similar analyses have also been carried out for black holes arising in alternative (2+1)-dimensional gravity theories~\cite{Becar:2019hwk, Canate:2019sbz, Anabalon:2019zae, Ovgun:2018gwt, Rincon:2018sgd, Gonzalez:2014voa, Catalan:2014una, Chen:2009hg, Chakrabarti:2009ww, Fernando:2009tv, Fernando:2008hb, Lopez-Ortega:2006tjo}. However, when higher-derivative corrections are introduced, as in the case of regular BTZ solutions derived from infinite Lovelock towers, the resulting perturbation equations typically become too complicated for analytic treatment. In such scenarios, numerical approaches, like the \emph{Leaver's continued fraction method}~\cite{Leaver:1990zz, Leaver:1985ax, Daghigh:2022uws} or the \emph{Horowitz–Hubeny method}~\cite{Horowitz:1999jd} become indispensable for determining the QNM spectrum.

In this work, we undertake a detailed study of scalar field perturbations in the background of the recently constructed regular BTZ black holes arising from an infinite tower of regularized Lovelock corrections. We derive the radial perturbation equation and analyze its structure, demonstrating that it contains too many singular points to be reducible to a standard hypergeometric or Heun-type equation. Consequently, we apply numerical techniques, specifically the Leaver and Horowitz–Hubeny methods, to compute the quasinormal frequencies and analyze their dependence on the regularization parameter ($\ell$). The results show strong agreement between the two methods, confirming the robustness of the QNM spectrum in these novel geometries and providing further evidence for their dynamical stability.

Our findings extend the landscape of known QNM spectra in three-dimensional gravity and offer new insights into the impact of infinite higher-curvature corrections on the ringdown dynamics of regular black holes. Given the increasing precision of gravitational wave observations, understanding how such corrections modify the QNM spectrum is not only of theoretical interest but may eventually yield observational signatures of ultraviolet modifications to general relativity.

Recent studies have shown that black hole quasinormal spectra can exhibit a remarkably rich “phase structure’’ when one moves in parameter space, with modes reorganizing themselves through bifurcations and branchings rather than forming a single simple ladder of damped oscillations. A paradigmatic example is provided by nearly extremal Kerr black holes, where the QNM spectrum splits into two distinct families, the so-called zero–damped modes (ZDMs) and damped modes (DMs): as the spin approaches its extremal value, part of the spectrum bifurcates into a branch of long-lived modes whose imaginary part tends to zero, while another branch retains a finite decay rate. This phenomenon was first identified in the branching analysis of nearly extremal Kerr QNMs and then explored in detail in terms of spectrum bifurcation and power-law ringdown.\cite{Yang:2012pj,Yang:2013uba,Zimmerman:2015rua} Similar ZDM/DM splitting and mode branching has since been observed for more general rotating/charged, nearly extremal geometries and in Kerr-Sen black holes~\cite{Konoplya:2013rxa, Mark:2014aja, Kokkotas:2015uma, Eniceicu:2019npi}. A different but related type of bifurcation arises for Maxwell perturbations of asymptotically AdS black holes and black branes: as the horizon radius or wavenumber is varied, a single complex-frequency branch can split into two purely imaginary branches, with an intricate re-labelling of overtones and Feigenbaum-like behavior in the asymptotic spectrum~\cite{Daghigh:2022uws, Morgan:2013dv, Wang:2015goa, Wang:2021upj, Lei:2021kqv, Fortuna:2022fdd}. This mechanism has been traced back both in black-hole backgrounds and in controlled toy models such as modified Pöschl–Teller potentials~\cite{Li:2024npg}. Scalar perturbations of Coulomb–like AdS black holes in $2+1$ dimensions provide yet another example of a QNM spectrum that splits into complex and purely imaginary branches as the charge is increased~\cite{Aragon:2021ogo}. In this work we show that the scalar QNM spectrum of regular BTZ black holes constructed from an infinite Lovelock tower displays an analogous bifurcation pattern, controlled by the regularization scale $\ell$ and the harmonic index $m$, thereby adding a new and qualitatively different (three–dimensional, regular) example to this emerging family of bifurcating QNM spectra.

The remainder of this paper is organized as follows. In Sec.~\ref{sec2} we review the construction of regular BTZ black holes from an infinite Lovelock tower. In Sec.~\ref{sec3} we derive the scalar perturbation equation and the associated effective potential. In Secs.~\ref{sec4} and \ref{sec5} we apply Leaver’s continued fraction method and the Horowitz–Hubeny method, respectively, to compute the QNM spectrum. Our numerical results and their interpretation in terms of bifurcation and mode branching are presented in Sec.~\ref{sec6}, and we conclude in Sec.~\ref{sec7}. Technical details, including the explicit recurrence coefficients, the Gaussian–elimination reduction of the eight–term relation, and the BTZ benchmark implementation of Leaver’s method, are collected in the Appendices \ref{app1} and \ref{app2}.

\section{Regular BTZ black holes from an infinite tower of corrections}\label{sec2}

In this section, we review the construction of regular BTZ black hole solutions in \(2+1\) dimensions, following Ref.~\cite{Fernandes:2025eoc}. For detailed derivations, we refer the reader to sections II and III of that work and the references therein.

The key idea is that the same \emph{infinite tower of higher-curvature densities} that regularizes the Schwarzschild singularity in quasi-topological gravity can be adapted to three dimensions, remarkably, without introducing Ostrogradsky instabilities. In the \(2+1\)-dimensional context, this tower is resummed into a shift-symmetric Horndeski theory governed by the action
\begin{equation}
S=\int d^{3}x\;\sqrt{-g}\left[R - 2\Lambda
+ \frac{1}{\ell^{2}} \sum_{n=2}^{\infty} c_{n} \ell^{2n} \mathcal{L}^{(n)} \right].
\end{equation}
Here, \(\Lambda\) is the cosmological constant, related to the bare AdS radius by \(\Lambda = -1/L^2\); \(\ell\) denotes the higher-derivative scale; and the coefficients \(c_n = \mathcal{O}(1)\) ensure a finite radius of convergence. This construction mirrors the higher-dimensional quasi-topological framework that cures curvature singularities in the Schwarzschild case~\cite{Bueno:2024dgm}, and provides the foundation for our quasinormal-mode analysis.

The terms \(\mathcal{L}^{(n)}\) are obtained via a dimensional regularization of the \(n\)th Lovelock invariant~\cite{Lovelock:1971yv},
\[
R^{(n)} = \frac{1}{2^{n}} \delta^{\mu_{1}\nu_{1}\dots\mu_{n}\nu_{n}}_{\alpha_{1}\beta_{1}\dots\alpha_{n}\beta_{n}}
\prod_{i=1}^{n} R^{\alpha_{i}\beta_{i}}_{\phantom{\alpha_{i}\beta_{i}} \mu_{i}\nu_{i}}.
\]
Specifically, the dimensionally regularized Lagrangian is defined as
\begin{equation}
\mathcal{L}^{(n)} = \lim_{d \to 2n} \frac{ \sqrt{-g}\,R^{(n)} - \sqrt{-\tilde{g}}\,\tilde{R}^{(n)} }{d - 2n}
\end{equation}
where \(\tilde{g}_{\mu\nu} = e^{-2\varphi} g_{\mu\nu}\) is a Weyl-related auxiliary metric. This limiting procedure yields, in three dimensions, a Horndeski-type scalar-tensor Lagrangian:
\begin{equation}
\begin{aligned}
\mathcal{L}^{(n)} &= G^{(n)}_2(\varphi,X) - G^{(n)}_3(\varphi,X) \Box\varphi + G^{(n)}_4(\varphi,X) R \\
&\quad + G^{(n)}_{4X} \left[ (\Box\varphi)^2 - \nabla_\mu\nabla_\nu\varphi\,\nabla^\mu\nabla^\nu\varphi \right],
\end{aligned}
\end{equation}
where \(X = -\tfrac{1}{2} \nabla_\mu \varphi \nabla^\mu \varphi\), and the coupling functions are given by
\begin{equation}
\begin{aligned}
G^{(n)}_2 &= -2^{n+1} (n-1) X^n, \\
G^{(n)}_3 &= 2^n n X^{n-1}, \\
G^{(n)}_4 &= -\frac{2^{n-1} n}{2n - 3} X^{n-1}.
\end{aligned}
\end{equation}
A crucial property of this construction is that all field equations remain second-order when evaluated on circularly symmetric backgrounds.

Assuming the scalar field profile \(\varphi = \ln(r / r_0)\) automatically solves the scalar equation, and the gravitational field equations reduce to a single algebraic relation for the metric function \(f(r) \equiv f\):
\begin{equation}
\begin{aligned}
\frac{1}{\ell^{2}} \sum_{n=1}^{\infty} c_n \left[ -\frac{f}{r^2} \right]^n &= -\frac{f_{\text{BTZ}}}{r^2}, \\
f_{\text{BTZ}} &= -M + \frac{r^2}{L^2},
\end{aligned}
\label{feqn}
\end{equation}
with \(c_1 = 1\). Here, \(f_{\text{BTZ}}\) is the metric function for the static (non-rotating) BTZ solution. Truncating the series at any finite order \(n_{\max}\) yields a power-law divergence in \(f\) near the origin. Thus, to ensure regularity at \(r = 0\), one must take \(n_{\max} \to \infty\) and demand that \(\lim_{n\to\infty} c_n^{1/n} < \infty\). This makes the infinite series not merely optional, but essential for curing the BTZ core.

For the simplest case \(c_n = 1\), Eq.~\eqref{feqn} can be inverted analytically to yield the closed-form metric function
\begin{equation}
f = \frac{r^2 f_{\text{BTZ}}}{r^2 - \ell^2 f_{\text{BTZ}}},
\label{metric}
\end{equation}
so that the line element
\begin{equation}
ds^2 = -f\,dt^2 + \frac{dr^2}{f} + r^2 d\phi^2
\end{equation}
remains smooth at the origin. In the \(r \to 0\) limit, the metric function behaves as \(f \sim -r^2/\ell^2\). The outer event horizon lies at \(r_\text{h} = L \sqrt{M}\), just as in the BTZ solution, while an extremal inner horizon appears at \(r = 0\), thereby avoiding the classical mass-inflation instability. To make the regularity of the geometry explicit, we evaluate the curvature invariants. The Ricci and Kretschmann scalars are
\begin{equation}
\begin{split}
R& =-\frac{2 f'}{r}-f'',\\
K& \equiv R_{\mu\nu\rho\sigma}R^{\mu\nu\rho\sigma}=\frac{2(f')^2}{r^2}+(f'')^2.
\end{split}
\end{equation}
At the origin, one finds 
\begin{equation}
R|_{r=0}=6/\ell^{2} \quad \text{and} \quad K|_{r=0}=12/\ell^{4},
\end{equation}
confirming that the spacetime is regular there. These invariants are also finite at the horizon \(r_\text{h}\):
\begin{equation}
\begin{split}
R\big|_{r=r_\text{h}}&=-\frac{8\ell^2+6L^2}{L^4},\\
K\big|_{r=r_\text{h}}&=\frac{4\left(16\ell^4+8\ell^2L^2+3L^4\right)}{L^8}.
\end{split}
\end{equation}

At large distances, the geometry asymptotes to \(\mathrm{AdS}_3\) with an effective AdS radius
\begin{equation}\label{leff}
L_{\text{eff}}^2 = L^2 - \ell^2.
\end{equation}
Throughout we restrict to $0\le \ell < L$ (i.e. $\ell<1$ in units $L=1$), so that $L_{\rm eff}^2=L^2-\ell^2>0$ and the geometry remains asymptotically AdS$_3$.
This provides a well-behaved and physically motivated background in which we now proceed to analyze the scalar QNM spectrum.

\section{Scalar Perturbations}\label{sec3}

In this section, we derive the equations governing massless scalar field perturbations in the spacetime described by the regular BTZ metric function given in \cref{metric}. The dynamics of the scalar field are governed by the generally covariant Klein–Gordon equation:
\begin{equation}
\label{KGE}
\Box \Phi = \frac{1}{\sqrt{-g}} \partial_{\mu} \left( \sqrt{-g} \, g^{\mu\nu} \partial_{\nu} \Phi \right) = 0.
\end{equation}
To separate variables, we adopt the ansatz
\begin{equation}
\Phi(t,r,\phi) = e^{-i\omega t} e^{i m \phi} R(r),
\end{equation}
where \(m = 0,1,2,\dots\) denotes the angular momentum quantum number. Substituting into \cref{KGE} leads to a second-order differential equation for the radial function \(R(r)\equiv R\):
\begin{equation} \label{KGradial}
\frac{d}{dr}\left( r f \frac{dR}{dr} \right) + \left( \frac{r \omega^2}{f} - \frac{m^2}{r} \right) R = 0.
\end{equation}
This equation features six singular points located at \(r_0 = 0\), \(r_{\text{h}}^{\pm} = \pm L\sqrt{M}\), \(r_{\ell}^{\pm} =  \pm i \ell L \sqrt{M} / L_\text{eff}\), and \(r_\infty = \infty \). The presence of these singularities prevents the transformation of \cref{KGradial} into a standard hypergeometric or Heun differential equation~\cite{Fontana:2022whx, Valtancoli:2016krb, Hortacsu:2011rr, Suzuki:1998vy, Suzuki:1999nn, Hatsuda:2020sbn}. In contrast to the BTZ case~\cite{Cardoso:2001hn, Birmingham:2001hc, Gupta:2015uga}, where such analytical methods are often applicable, the complexity here necessitates a numerical treatment.

To this end, we compute the QNM frequencies using two independent numerical techniques:

\begin{itemize}
\item The \emph{Leaver's continued fraction method}~\cite{Leaver:1990zz, Leaver:1985ax, Livine:2024bvo}, which is among the most precise techniques available for determining QNMs.
\item The \emph{Horowitz–Hubeny method}~\cite{Horowitz:1999jd}, a well-established approach in AdS backgrounds (see also Refs.~\cite{Cardoso:2001hn, Cardoso:2001bb, DalBoscoFontana:2023syy, Aragon:2021ogo}).
\end{itemize}

To further analyze the perturbation equation, it is convenient to recast \cref{KGradial} into a Schrödinger-like form. This is achieved by the field redefinition \( R(r) = \chi(r)/\sqrt{r} \) and introducing the tortoise coordinate \( r_* \), defined via \( dr_* = dr/f \). The equation then becomes:
\begin{equation} \label{waveeqn}
\frac{d^2 \chi}{dr_*^2} + \left[ \omega^2 - V_\text{eff}(r) \right] \chi = 0,
\end{equation}
where the effective potential \(V_\text{eff}(r)\) is given by
\begin{widetext}
\begin{eqnarray}
V_\text{eff}(r) &=& \frac{f}{4r^2} \left[ 2r f' - f + 4m^2 \right] \nonumber \\
&=& -\frac{\left(L^2 M - r^2\right)}{4 \left( \ell^2 L^2 M + r^2 (L^2 - \ell^2) \right)^3}
\Bigg[ \ell^2 r^2 (L^2 M - r^2) \left( L^2 (8m^2 - 3M) + 3r^2 \right) 
+ 4\ell^4 m^2 (r^2 - L^2 M)^2 \nonumber \\
&& \qquad \qquad + L^4 r^4 (4m^2 + M) + 3L^2 r^6 \Bigg]. \label{veff}
\end{eqnarray}
\end{widetext}

\begin{figure}[t!]
\centering
\includegraphics[width=0.45\textwidth]{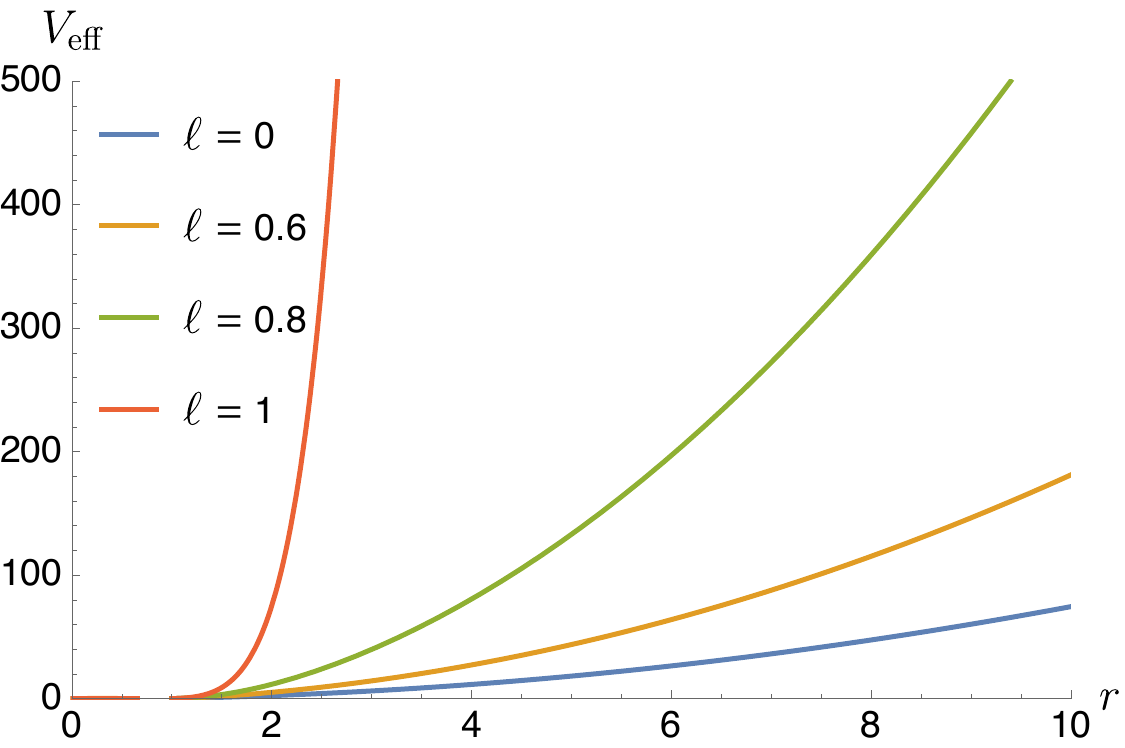}
\caption{Effective potential for the massless scalar field for different values of the regularization parameter \(\ell\), shown for the fundamental mode \(m=0\). Parameters are fixed as \(M = 1\), \(L = 1\).}
\label{fig1}
\end{figure}

The qualitative behavior of \(V_\text{eff}(r)\) is illustrated in \cref{fig1} for various values of the regularization parameter \(\ell\). As expected in AdS spacetimes, the potential remains positive outside the horizon and diverges at spatial infinity. Notably, increasing \(\ell\) amplifies the potential barrier across the entire exterior region.

\section{Leaver's Method}\label{sec4}

We now compute the QNM frequencies using the continued fraction method introduced by Leaver~\cite{Leaver:1985ax, Leaver:1990zz}. While this method has been successfully applied to asymptotically AdS black holes in four dimensions~\cite{Daghigh:2022uws} (see also~\cite{Moss:2001ga, Berti:2003ud}),  to the best of our knowledge, there exist only very few applications of the Leaver's method in $(2+1)$ dimensions~\cite{Cardoso:2004fi,Matyjasek:2024uwo}, especially the BTZ-like spacetimes. In this work, we extend its application to the regular BTZ black holes discussed earlier. For completeness, we also provide its implementation in the case of the standard BTZ solution in Appendix~\ref{app2}, which serves as a consistency check.

In asymptotically AdS spacetimes, the appropriate boundary conditions for QNMs require that the waves are purely ingoing at the event horizon and vanish at spatial infinity. That is,
\begin{equation}\label{bc1}
\chi(r) \sim e^{-i \omega r_*} \quad \text{as} \quad r \rightarrow r_\text{h},
\end{equation}
\begin{equation}\label{bc2}
\chi(r) \sim 0 \quad \text{as} \quad r \rightarrow \infty.
\end{equation}
To analyze the asymptotic behavior, we rewrite the Schrödinger-like equation \cref{waveeqn} in terms of the radial coordinate:
\begin{equation}
f \frac{d^2 \chi}{dr^2} + f' \frac{d\chi}{dr} + \left[ \frac{\rho^2}{f} - \frac{V_\text{eff}}{f} \right] \chi = 0,
\label{Waveradial}
\end{equation}
where \(\rho = -i \omega\). From this, and using the boundary conditions in \cref{bc1}-\cref{bc2}, we find the asymptotic behavior of the wave function:
\begin{equation}
\chi(r) \xrightarrow{r \rightarrow r_\text{h}} (r - r_\text{h})^{b \rho}, \qquad
\chi(r) \xrightarrow{r \rightarrow \infty} r^{-3/2},
\label{scalarbc}
\end{equation}
with
\begin{equation}
b = \frac{\left[\ell^2 r_\text{h}^2 - L^2 \left( \frac{\ell^2 r_\text{h}^2}{L^2} + r_\text{h}^2 \right)\right]^2}{2 L^2 r_\text{h}^5}.
\label{bvalue}
\end{equation}

A solution satisfying the above asymptotic structure can be written as a Frobenius-like series:
\begin{equation}
\chi(r) = \left( \frac{r - r_\text{h}}{r} \right)^{b \rho} \left( \frac{r_\text{h}}{r} \right)^{3/2} \sum_{n=0}^{\infty} a_n z^n,
\label{ansatz}
\end{equation}
where $z=(r - r_\text{h})/r$. The above ansatz encodes the quasinormal boundary conditions, where the full solution is represented as the product of a singular factor (capturing the asymptotic behavior) and a series expansion that remains convergent within the domain \( r_\text{h} \leq r < \infty \). For the series to converge at spatial infinity \( r = \infty \), which corresponds to \( z = 1 \) under the conformal mapping used in the Leaver method, all singular points of the transformed radial equation must lie outside the unit circle in the complex \( z \)-plane. In our case, the singular point \( r_\text{h}^- \) always lies outside the unit circle, but the two additional singularities \( r_\ell^\pm \), arising from the regularization structure, satisfy \( |z| > 1 \) when \( \ell < 1 \), thereby lying outside the unit circle. Accordingly, in what follows, we restrict our analysis of QNMs to the regime $\ell <1$. This is not a purely numerical choice: it coincides with the geometric requirement \, $L_{\text{eff}}^2 = L^2 - \ell^2 > 0$\, from eq.~\ref{leff}, so that in the
units $L = 1$ used throughout the numerical analysis the condition $\ell < L$ simply becomes $\ell < 1$. For \( \ell \geq 1 \), the Frobenius series may fail to converge at infinity unless an analytic continuation is performed through appropriate midpoints. 

Substituting the ansatz~\eqref{ansatz} into the radial equation~\eqref{Waveradial} yields an eight-term recurrence relation of the form:
\begin{widetext}
\begin{eqnarray}
\alpha_1 a_1 + \beta_1 a_0 &=& 0, \\
\alpha_2 a_2 + \beta_2 a_1 + \gamma_2 a_0 &=& 0, \\
\alpha_3 a_3 + \beta_3 a_2 + \gamma_3 a_1 + \delta_3 a_0 &=& 0, \\
\alpha_4 a_4 + \beta_4 a_3 + \gamma_4 a_2 + \delta_4 a_1 + \zeta_4 a_0 &=& 0, \\
\alpha_5 a_5 + \beta_5 a_4 + \gamma_5 a_3 + \delta_5 a_2 + \zeta_5 a_1 + \eta_5 a_0 &=& 0, \\
\alpha_6 a_6 + \beta_6 a_5 + \gamma_6 a_4 + \delta_6 a_3 + \zeta_6 a_2 + \eta_6 a_1 + \theta_6 a_0 &=& 0, \\
\alpha_n a_n + \beta_n a_{n-1} + \gamma_n a_{n-2} + \delta_n a_{n-3} + \zeta_n a_{n-4} + \eta_n a_{n-5} + \theta_n a_{n-6} + \tau_n a_{n-7} &=& 0, \quad n \geq 7.
\label{sevenrecurrence}
\end{eqnarray}
\end{widetext}
Here, \(a_0\) can be taken as a normalization constant (e.g., \(a_0 = 1\)). The recurrence coefficients are detailed in Appendix~\ref{app1}. To reduce this high-order recurrence to a more manageable form, we use Gaussian elimination to obtain a three-term recurrence relation:
\begin{equation}
\alpha_n^{(5)} a_n + \beta_n^{(5)} a_{n-1} + \gamma_n^{(5)} a_{n-2} = 0.
\label{eq-threeterm}
\end{equation}
Because this reduction is recursive, closed-form expressions for \(\alpha_n^{(5)}\), \(\beta_n^{(5)}\), and \(\gamma_n^{(5)}\) are not available in general and must be computed numerically.

\begin{figure*}[t!]
\centering
\includegraphics[width=0.45\textwidth]{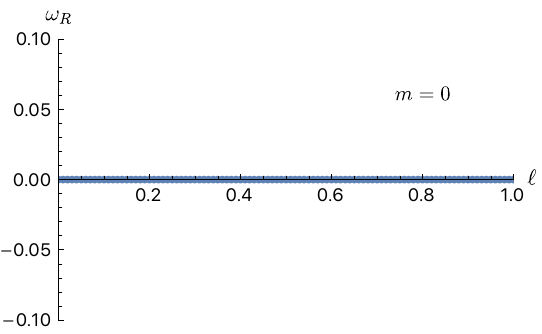}
\includegraphics[width=0.45\textwidth]{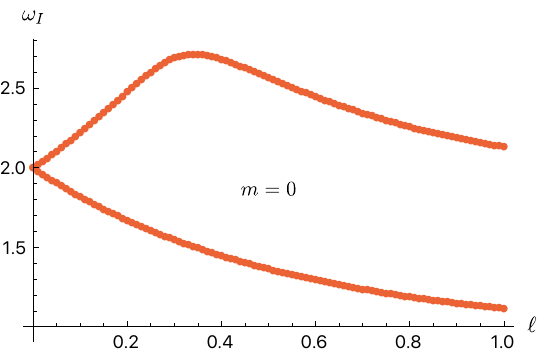}
\includegraphics[width=0.45\textwidth]{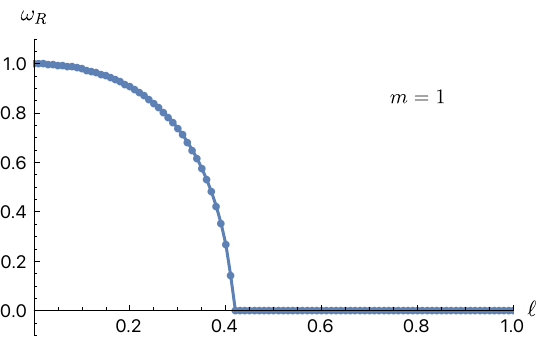}
\includegraphics[width=0.45\textwidth]{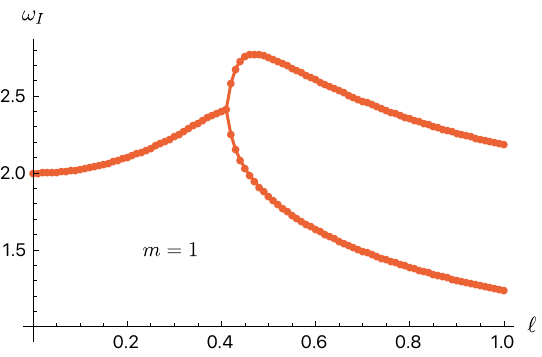}
\includegraphics[width=0.45\textwidth]{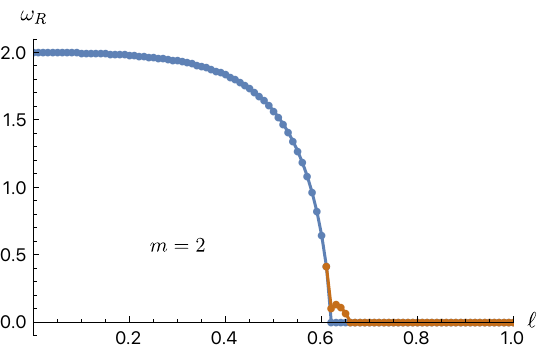}
\includegraphics[width=0.45\textwidth]{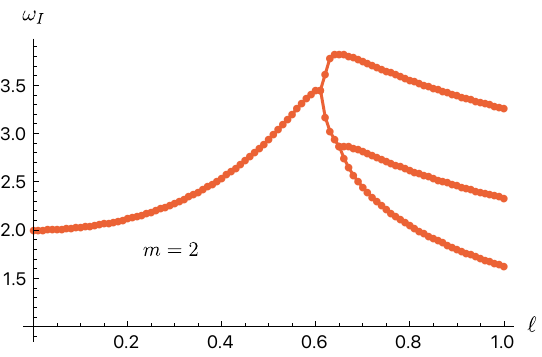}
\caption{QNM frequencies for the first three harmonics $m =0,1,2$ in the range \(0 \leq \ell < 1\). The left panel shows the dependence of the real part \(\omega_R\) on \(\ell\), while the right panel shows the corresponding imaginary part \(\omega_I\). For \(m = 0\), the BTZ black hole has a vanishing real part \(\omega_R = 0\), so the splitting of \(\omega_I\) into two branches, the upper (more damped) and the lower (less damped), occurs already at \(\ell = 0\). For \(m = 1\), the BTZ value is \(\omega_R = 1\); as \(\ell\) increases, \(\omega_R\) decreases and crosses zero at \(\ell = \ell_{\text{crit}} = 0.433013\), where \(\omega_I\) splits into two branches. For \(m = 2\), the BTZ value is \(\omega_R = 2\); \(\omega_R\) decreases with \(\ell\) and becomes zero at \(\ell = \ell_{\text{crit}} = 0.654654\), and \(\omega_I\) again splits into two branches. In this case \(\omega_R\) also exhibits a small jump, leading to a short interval in which a nonzero \(\omega_R\) branch coexists with the zero branch; when this jump merges back into the zero branch, the corresponding \(\omega_I\) spectrum develops a third branch.}
\label{fig2}
\end{figure*}

\begin{figure*}[t!]
\centering
\includegraphics[width=0.45\textwidth]{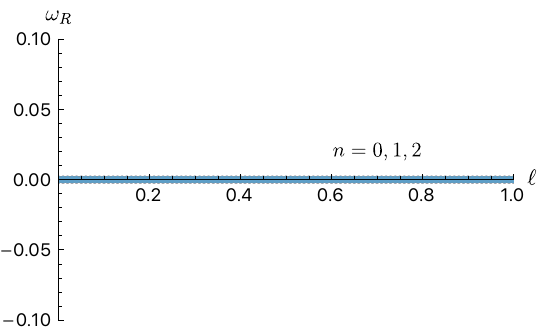}
\includegraphics[width=0.45\textwidth]{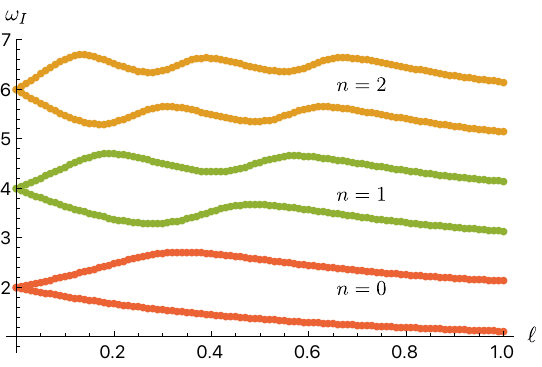}
\includegraphics[width=0.45\textwidth]{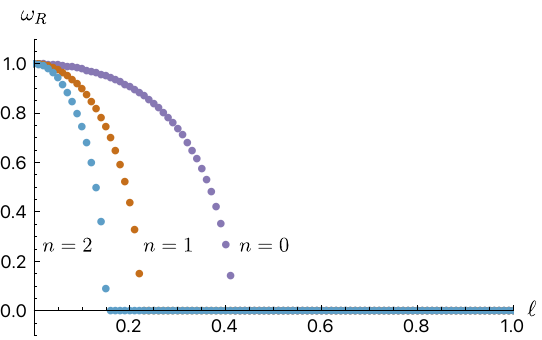}
\includegraphics[width=0.45\textwidth]{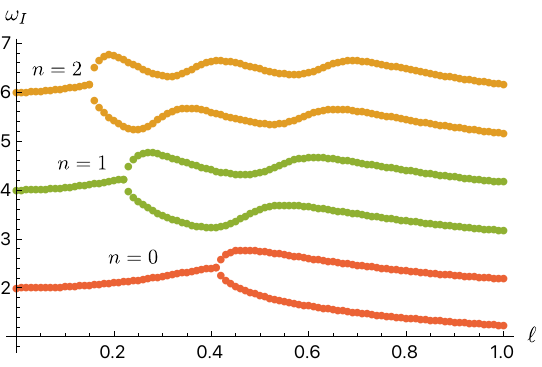}
\includegraphics[width=0.45\textwidth]{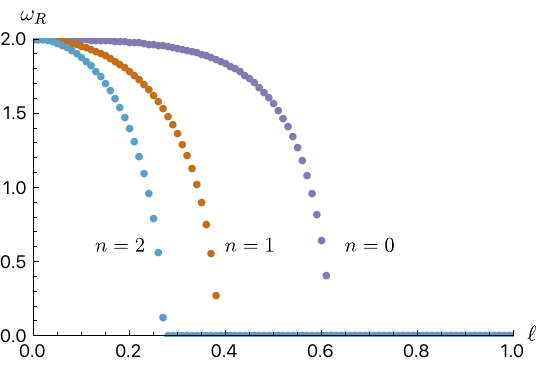}
\includegraphics[width=0.45\textwidth]{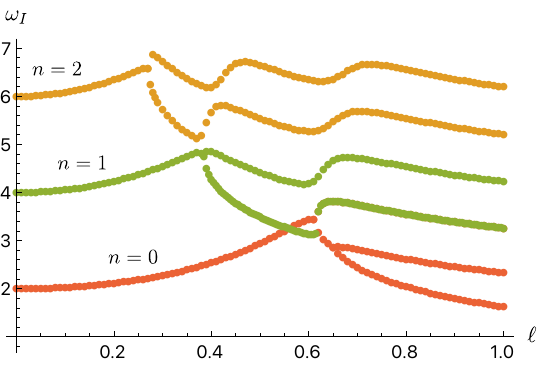}
\caption{Fundamental mode $(n=0)$ and the first two overtones $(n=1,2)$ of the scalar QNM frequencies as functions of the regularization parameter in the range \(0 \leq \ell < 1\). The left panels display the dependence of the real part \(\omega_R\) on \(\ell\), while the right panels show the corresponding imaginary part \(\omega_I\). For \(m = 0\) all modes are purely imaginary and the regular BTZ geometry splits the BTZ value at \(\ell = 0\) into two branches with different damping rates, the upper (more damped) and the lower (less damped). For \(m = 1\) and \(m = 2\), the BTZ-like complex branches move towards the imaginary axis as \(\ell\) increases and, at critical values of \(\ell\), bifurcate into multiple purely imaginary branches. In the \(m = 2\) case, this bifurcation pattern is particularly pronounced and leads to a nontrivial reordering of the fundamental mode and its first overtones, as evidenced by the crossing of the corresponding curves.}
\label{fig3}
\end{figure*}

The convergence of the Frobenius series \eqref{ansatz} is governed by a continued fraction relation~\cite{Leaver:1985ax}:
\begin{eqnarray}
\beta_1^{(5)} &=& \frac{\alpha_1^{(5)} \gamma_2^{(5)}}{\beta_2^{(5)} - \frac{\alpha_2^{(5)} \gamma_3^{(5)}}{\beta_3^{(5)} - \cdots}} \nonumber \\
&=& \frac{\alpha_1^{(5)} \gamma_2^{(5)}}{\beta_2^{(5)} -} \frac{\alpha_2^{(5)} \gamma_3^{(5)}}{\beta_3^{(5)} -} \frac{\alpha_3^{(5)} \gamma_4^{(5)}}{\beta_4^{(5)} -} \cdots
\label{eqLeaver}
\end{eqnarray}
The QNM frequencies are determined as the roots of this continued fraction equation. In practice, the series is truncated at a large but finite order \(n\), and the tail end is approximated using the Nollert improvement technique~\cite{Nollert:1993zz}. The resulting frequencies for various $m$ and $\ell$ values are shown in Figs.~\ref{fig2}, \ref{fig3}, and Table~\ref{tab1}. We work in units where \(L = 1\) and \( M=1\) throughout the article. Before analyzing the results, we present the Horowitz–Hubeny method for calculating the QNM spectrum in order to ensure the validity and robustness of our findings.

\begin{table*}[t]
\centering
\vspace{0.5cm}
\caption{QNM frequencies with \(M=1\) and \(L=1\). For each angular index
\(m\), rows list the overtone number \(n\), while columns list the
regularization parameter \(\ell\). When multiple rows with the same \(n\)
appear within a given \(m\)-block, they correspond to distinct branches
of the same overtone arising from the bifurcation of the spectrum. Leaver and
Horowitz--Hubeny methods agree up to three decimal places for all displayed
entries, so only one numerical value is shown for each QNM branch. All
frequencies are rounded to three decimals.}
\label{tab1}
\begin{tabular}{c|@{\hspace{10pt}}c@{\hspace{10pt}}c@{\hspace{10pt}}c@{\hspace{10pt}}c@{\hspace{10pt}}c@{\hspace{10pt}}c}
\hline\hline
$n \,\backslash\, \ell$ & $0$ & $0.2$ & $0.4$ & $0.6$ & $0.8$ & $0.99$ \\
\hline
\multicolumn{7}{c}{$m = 0$} \\
\hline
$0$
  & $0.000 - 2.000\,i$
  & $0.000 - 1.671\,i$
  & $0.000 - 1.449\,i$
  & $0.000 - 1.296\,i$
  & $0.000 - 1.190\,i$
  & $0.000 - 1.120\,i$
  \\
  &
  &
  $0.000 - 2.477\,i$
  & $0.000 - 2.682\,i$
  & $0.000 - 2.448\,i$
  & $0.000 - 2.258\,i$
  & $0.000 - 2.138\,i$
  \\[6pt]
$1$
  & $0.000 - 4.000\,i$
  & $0.000 - 3.381\,i$
  & $0.000 - 3.532\,i$
  & $0.000 - 3.570\,i$
  & $0.000 - 3.313\,i$
  & $0.000 - 3.147\,i$
 \\
  &
  &
  $0.000 - 4.703\,i$
  & $0.000 - 4.335\,i$
  & $0.000 - 4.647\,i$
  & $0.000 - 4.361\,i$
  & $0.000 - 4.152\,i$
 \\[6pt]
$2$
  & $0.000 - 6.001\,i$
  & $0.000 - 5.334\,i$
  & $0.000 - 5.494\,i$
  & $0.000 - 5.623\,i$
  & $0.000 - 5.406\,i$
  & $0.000 - 5.156\,i$
  \\
  &
  &
  $0.000 - 6.526\,i$
  & $0.000 - 6.633\,i$
  & $0.000 - 6.466\,i$
  & $0.000 - 6.448\,i$
  & $0.000 - 6.160\,i$
  \\
[6pt]
\hline
\multicolumn{7}{c}{$m = 1$} \\
\hline
$0$
  & $1.000 - 2.000\,i$
  & $0.907 - 2.103\,i$
  & $0.269 - 2.403\,i$
  & $0.000 - 1.634\,i$
  & $0.000 - 1.387\,i$
  & $0.000 - 1.243\,i$
  \\
  &
  &
  &
  &
  $0.000 - 2.607\,i$
  & $0.000 - 2.353\,i$
  & $0.000 - 2.192\,i$
\\[6pt]
$1$
  & $1.000 - 4.000\,i$
  & $0.439 - 4.191\,i$
  & $0.000 - 3.230\,i$
  & $0.000 - 3.655\,i$
  & $0.000 - 3.375\,i$
  & $0.000 - 3.180\,i$
\\
  &
  &
  &
  $0.000 - 4.413\,i$
  & $0.000 - 4.663\,i$
  & $0.000 - 4.407\,i$
  & $0.000 - 4.176\,i$
\\[6pt]
$2$
  & $1.000 - 6.001\,i$
  & $0.000 - 5.418\,i$
  & $0.000 - 5.594\,i$
  & $0.000 - 5.550\,i$
  & $0.000 - 5.441\,i$
  & $0.000 - 5.174\,i$
\\
  &
  &
  $0.000 - 6.750\,i$
  & $0.000 - 6.632\,i$
  & $0.000 - 6.398\,i$
  & $0.000 - 6.477\,i$
  & $0.000 - 6.174\,i$
\\[6pt]
\hline
\multicolumn{7}{c}{$m = 2$} \\
\hline
$0$
  & $2.000 - 2.000\,i$
  & $1.979 - 2.117\,i$
  & $1.834 - 2.539\,i$
  & $0.643 - 3.444\,i$
  & $0.000 - 2.047\,i$
  & $0.000 - 1.640\,i$
  \\
  &
  &
  &
  &
  & $0.000 - 2.619\,i$
  & $0.000 - 2.344\,i$
  \\
  &
  &
  &
  &
  & $0.000 - 3.555\,i$
  & $0.000 - 3.274\,i$
  \\[6pt]
$1$
  & $2.000 - 4.000\,i$
  & $1.782 - 4.233\,i$
  & $0.000 - 4.284\,i$
  & $0.000 - 3.138\,i$
  & $0.000 - 3.555\,i$
  & $0.000 - 3.274\,i$
 \\
  &
  &
  &
  $0.000 - 4.853\,i$
  & $0.000 - 4.188\,i$
  & $0.000 - 4.540\,i$
  & $0.000 - 4.243\,i$
 \\[6pt]
$2$
  & $2.000 - 6.000\,i$
  & $1.396 - 6.345\,i$
  & $0.000 - 5.674\,i$
  & $0.000 - 5.275\,i$
  & $0.000 - 5.225\,i$
  & $0.000 - 5.225\,i$
  \\
  &
  &
  &
  $0.000 - 6.189\,i$
  & $0.000 - 6.347\,i$
  & $0.000 - 6.560\,i$
  & $0.000 - 6.215\,i$
  \\
\hline\hline
\end{tabular}
\vspace{0.5cm}
\end{table*}

\begin{figure}[t!]
\centering
\includegraphics[width=0.45\textwidth]{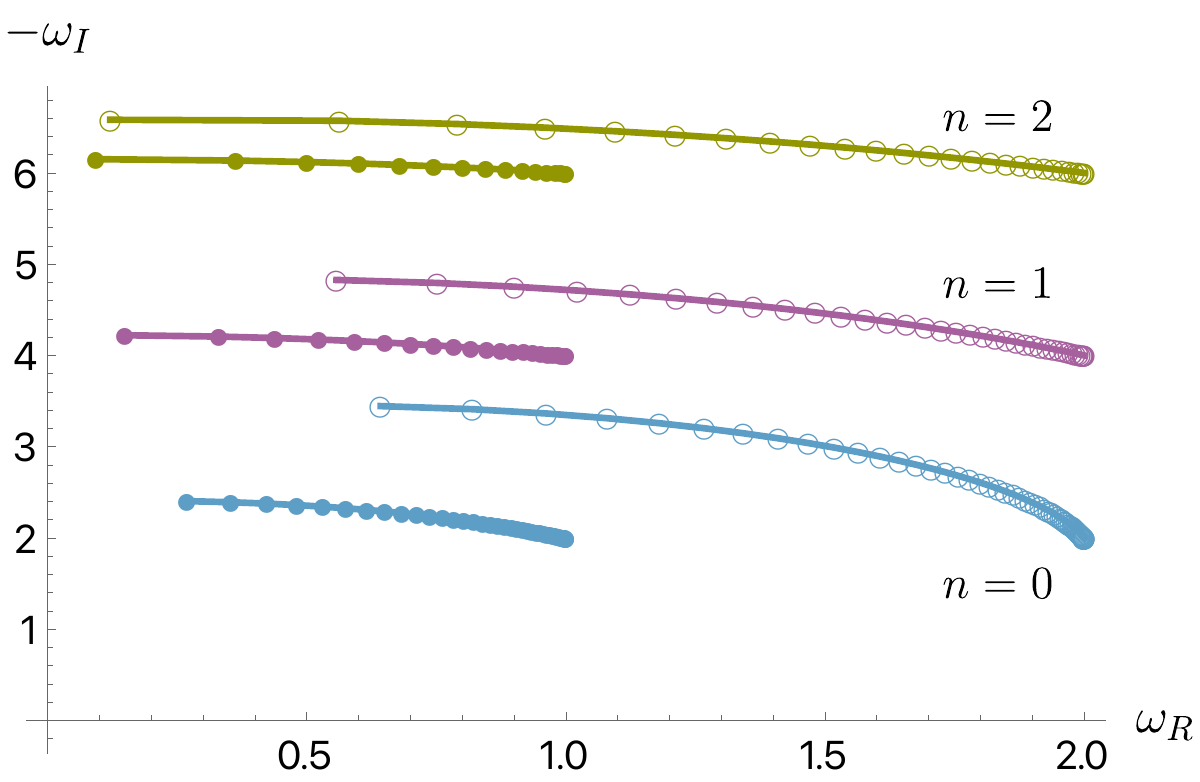}
\caption{The trajectory plot of the fundamental scalar quasinormal frequencies in the $(\omega_R,-\omega_I)$ plane for $m = 1$ and $m = 2$ as the regularization scale $\ell$ is varied. Solid circles (\(\bullet\)) denote the $m = 1$ modes, while open circles (\(\circ\)) correspond to $m = 2$.}
\label{fig4}
\end{figure}

\begin{figure*}[t!]
\centering
\includegraphics[width=0.45\textwidth]{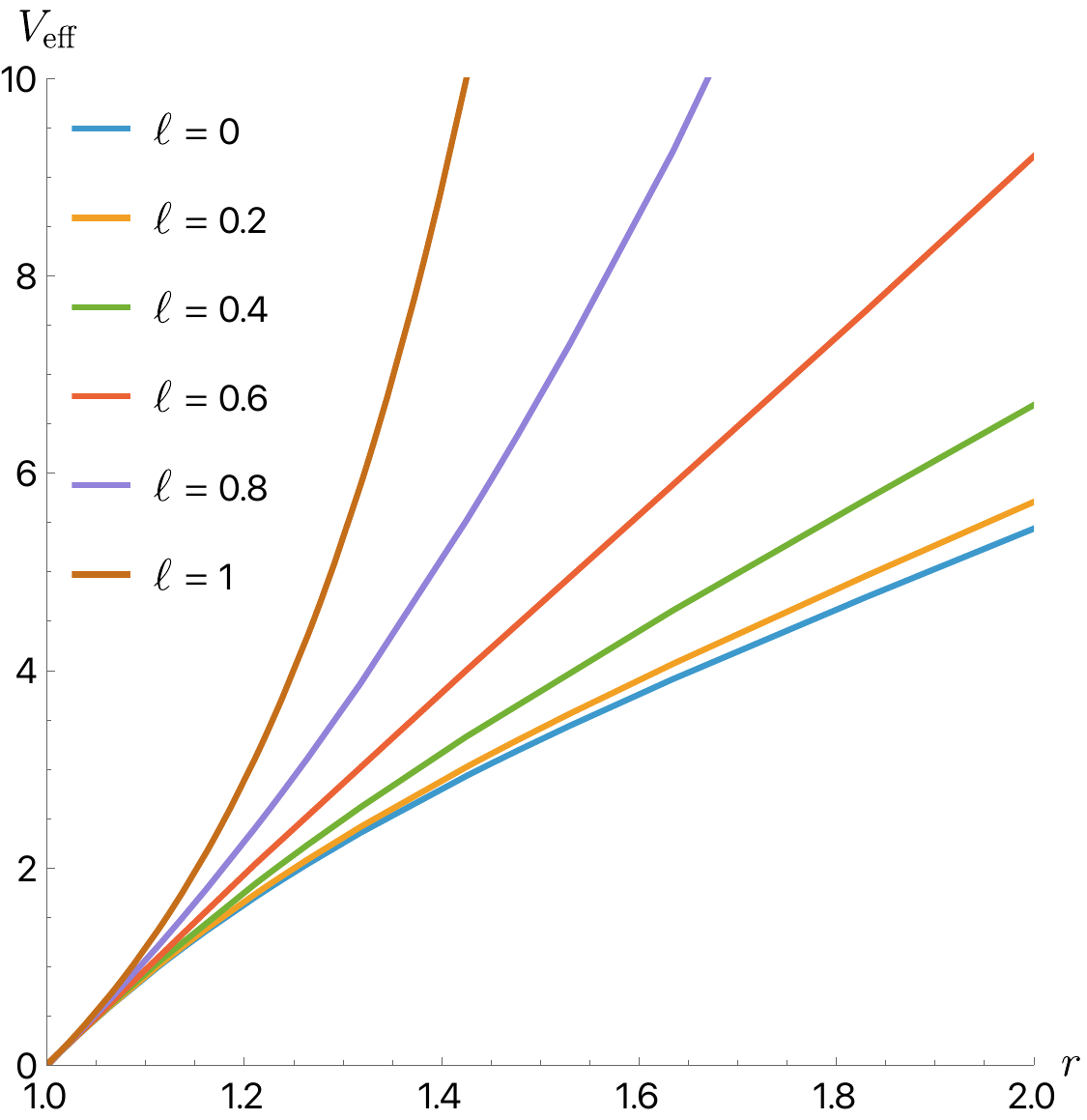}
\includegraphics[width=0.45\textwidth]{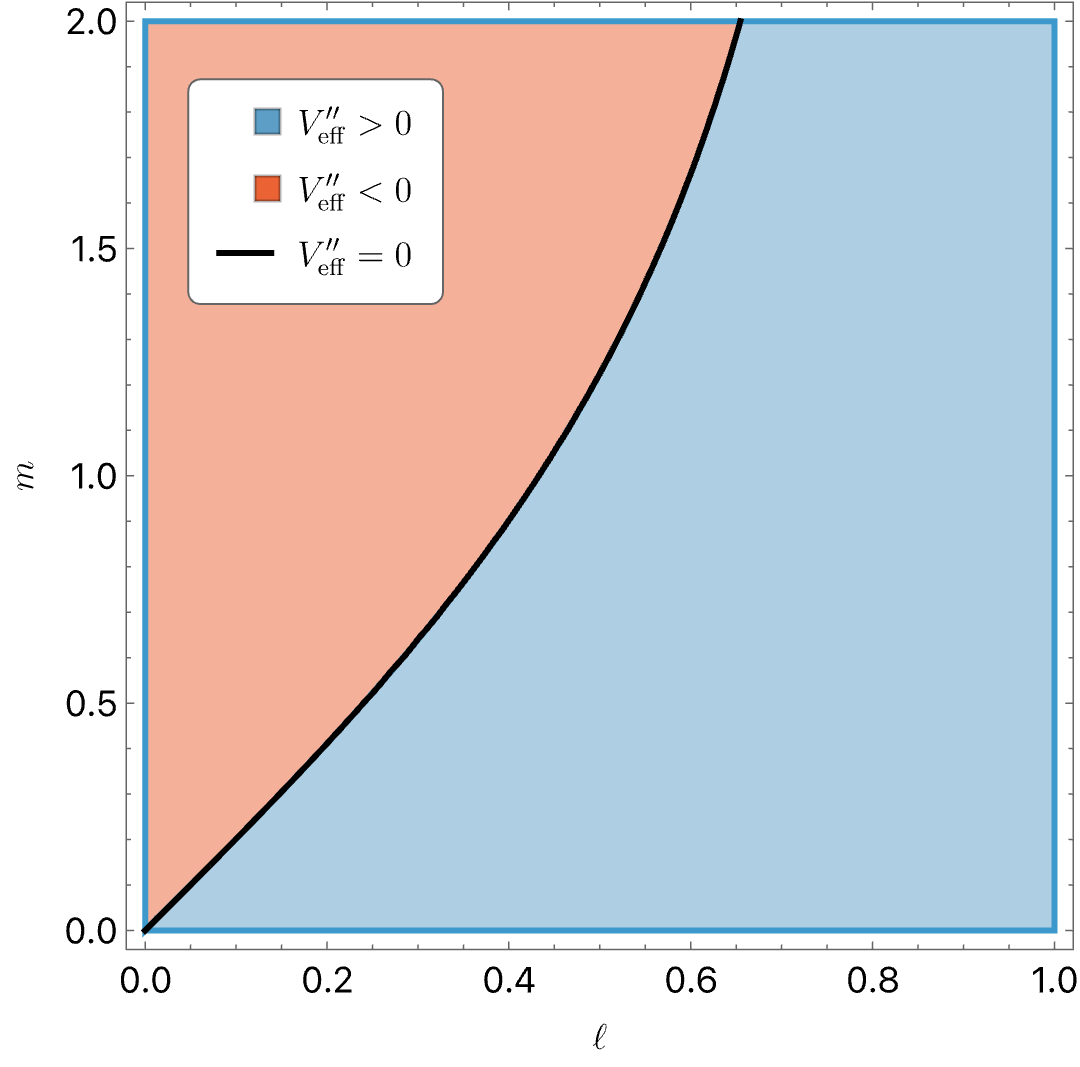}
\caption{Left: Near-horizon behavior of the effective potential for $M=L=1$ and $m=2$. The plot clearly shows that the concavity of the curve changes as $\ell$ is varied. Right: Contour plot of $V_{\text{eff}}''(r_\text{h})$ in the $(\ell, m )$–plane, showing the regions where the effective potential at the event horizon is concave up, concave down, or has a vanishing second derivative. The black line corresponds to the critical values $\ell_{\text{crit}}$ for the fundamental mode.}
\label{fig5}
\end{figure*}

\section{Horowitz--Hubeny Method}\label{sec5}

In this section, we compute the QNM frequencies using the Horowitz--Hubeny method~\cite{Horowitz:1999jd}, and use the results to independently confirm those obtained from Leaver's continued fraction approach.

To apply this method, we begin by rewriting the Schrödinger-like wave equation~\eqref{waveeqn} using the field redefinition \(\psi(r) = e^{i\omega r_*} \chi(r_*)\), resulting in:
\begin{equation}
\frac{d}{dr}\left(f\,\frac{d\psi}{dr}\right) - 2i\omega\,\frac{d\psi}{dr} - \frac{V_\text{eff}(r)}{f}\, \psi = 0\,.
\label{KGeq}
\end{equation}
We then introduce a new coordinate \(z = 1/r\), in which the event horizon is located at \(z_\text{h} = 1/r_\text{h}\). In terms of this coordinate, Eq.~\eqref{KGeq} becomes:
\begin{equation}
s(z)\,\frac{d^2 \psi}{dz^2} + \frac{t(z)}{z - z_\text{h}}\,\frac{d \psi}{dz} + \frac{u(z)}{(z - z_\text{h})^2}\, \psi = 0 \,,
\label{eq:HHform}
\end{equation}
where the functions \(s(z)\), \(t(z)\), and \(u(z)\) are given by:
\begin{align}
s(z) &= \frac{z^2 - L^2 M z^4}{\ell^2 \left(L^2 M z^2 - 1\right) + L^2} \,, \nonumber\\
t(z) &= 2 z^2 \left[-\frac{L^4 M z}{\left[\ell^2 \left(L^2 M z^2 - 1\right) + L^2\right]^2} + i\omega \right] \,, \nonumber\\
u(z) &= \frac{z^2 \left[\ell^2 \left(L^2 M z^2 - 1\right) + L^2\right]}{L^2 M z^2 - 1} \, V_\text{eff}(z)\,.
\label{eq:stu}
\end{align}

These are third-order polynomials in \(z\), and reduce to the standard BTZ case when \(\ell = 0\)~\cite{Cardoso:2001hn}. The functions are now expanded in a Taylor series about the horizon \(z = z_\text{h}\):
\begin{align}
s(z) &= \sum_{n=0}^{3} s_n\, (z - z_\text{h})^n \,, \\
t(z) &= \sum_{n=0}^{3} t_n\, (z - z_\text{h})^n \,, \\
u(z) &= \sum_{n=0}^{3} u_n\, (z - z_\text{h})^n \,.
\end{align}
Likewise, the wavefunction is expanded in a Frobenius series:
\begin{equation}
\psi(z) = (z - z_\text{h})^\alpha \sum_{n=0}^{\infty} a_n\, (z - z_\text{h})^n\,.
\label{eq:psiExpansion}
\end{equation}

To determine the exponent \(\alpha\), we substitute \(\psi(z) = (z - z_\text{h})^\alpha\) into Eq.~\eqref{eq:HHform}, yielding the indicial equation:
\begin{equation}
\alpha(\alpha - 1)\, s_0 + \alpha\, t_0 + u_0 = 0\,.
\label{eq:alpha}
\end{equation}
The physical boundary condition of purely ingoing waves at the horizon imposes \(\alpha = 0\).
Substituting the series expansions for \(s(z)\), \(t(z)\), \(u(z)\), and \(\psi(z)\) into Eq.~\eqref{eq:HHform}, we obtain the following recursive relation for the coefficients \(a_n\):
\begin{equation}
a_n = -\frac{1}{P_n} \sum_{j=0}^{n-1} \left[ j(j - 1)\, s_{n-j} + j\, t_{n-j} + u_{n-j} \right] a_j \,,
\label{eq:recursion}
\end{equation}
where
\begin{equation}
P_n = n(n - 1)\, s_0 + n\, t_0 + u_0\,.
\label{eq:Pn}
\end{equation}

To determine the QNM frequencies, we impose the Dirichlet boundary condition at spatial infinity (\(z \to 0\)), which requires:
\begin{equation}
\psi(z \to 0) = 0 \quad \Rightarrow \quad \sum_{n=0}^{\infty} a_n\, (-z_\text{h})^n = 0 \,.
\label{eq:Dirichlet}
\end{equation}
This boundary condition yields an equation for \(\omega\), which we solve numerically after truncating the series at a finite order \(n = N\). The roots of this equation provide the quasinormal frequencies. The resulting frequencies agree with those obtained using the Leaver method up to three decimal places and are shown in Figs.~\ref{fig2}, \ref{fig3}, and Table~\ref{tab1}.

\section{Results and Discussions}\label{sec6}
In order to interpret the QNM spectra, it is useful to recall that a generic quasinormal frequency can be written as \(\omega = \omega_R + i\,\omega_I\), where \(\omega_R\) sets the oscillation frequency and \(\omega_I<0\) governs the exponential decay of the ringdown. Modes with \(\omega_R\neq 0\) describe damped oscillations, whereas purely imaginary modes (\(\omega_R=0\)) correspond to overdamped, nonoscillatory relaxation of the late–time tail. For the BTZ black hole the scalar quasinormal spectrum is known to contain a purely imaginary family for the \(S\)-wave \((m=0)\), while the modes with \(m\neq 0\) are complex.\footnote{See, e.g., Refs.~\cite{Cardoso:2001hn,Birmingham:2001hc} for the BTZ scalar spectrum.}

Table~\ref{tab1} and Figs.~\ref{fig2}, \ref{fig3} show that the regular BTZ geometry qualitatively modifies this pattern while preserving stability: for all parameter values we considered, we find \(\omega_I<0\), so the spacetime remains linearly stable under scalar perturbations. For the dominant scalar mode (\(n=0\)) with \(m=0\), the BTZ black hole has a purely imaginary frequency, \(\omega_R=0\), and this property persists in the regular case. However, as soon as the regularity parameter \(\ell\) is turned on, the single BTZ value at \(\ell=0\) splits into two distinct purely imaginary branches in the regular BTZ spectrum (upper left and upper right panels of Fig.~\ref{fig2}). One of these branches has \(|\omega_I|\) larger than the BTZ value and is therefore more strongly damped, whereas the other branch has a smaller \(|\omega_I|\) and decays more slowly. 

For \(m>0\) the standard BTZ spectrum is purely complex. In the regular BTZ case, we again recover the BTZ frequencies at \(\ell=0\), but for \(\ell>0\) we find new families of modes that are purely imaginary in certain \(\ell\)-intervals. For \(m=1\), starting from the BTZ value at \(\ell=0\), the real part \(\omega_R\) decreases monotonically with increasing \(\ell\), while \(|\omega_I|\) initially grows. At a critical value \(\ell=\ell_\text{crit}= 0.433013\) the complex branch reaches the imaginary axis, \(\omega_R(\ell_\text{crit})=0\), and at this point the imaginary part \(\omega_I\) bifurcates into two purely imaginary modes (see the middle panels of Fig.~\ref{fig2}). For \(\ell>\ell_{\text{crit}}\) both branches remain purely imaginary: the upper branch is more strongly damped than the BTZ mode, while the lower branch has smaller \(|\omega_I|\) and governs the slowest late–time decay.

For \(m>1\) the spectrum exhibits an additional layer of structure. In the \(m=2\) case, the BTZ–like complex branch again moves towards the imaginary axis as \(\ell\) grows and becomes purely imaginary at a first critical point \(\ell=\ell_\text{crit}\), where \(\omega_I\) splits into two branches, as in the \(m=1\) case. For slightly larger \(\ell\), the real part \(\omega_R\) develops a short excursion away from zero, so that a nonzero-\(\omega_R\) branch briefly coexists with the zero–\(\omega_R\) branch. When this nonzero branch returns to the imaginary axis at a second critical value \(\ell=\ell_{\text{crit}}'\), the lower purely imaginary branch undergoes a secondary bifurcation. As a result, for \(\ell>\ell_{\text{crit}}'\) the spectrum contains three distinct purely imaginary branches (see the lower panels of Fig.~\ref{fig2} and the \(m=2\) entries in Table~\ref{tab1}). In all cases, the branch with the smallest magnitude \(|\omega_I|\) is the least damped and therefore dominates the very late–time tail of the scalar perturbation, while the more strongly damped branches are expected to contribute mainly to the early ringdown. The corresponding bifurcation patterns for the first two harmonics are displayed in Fig.~\ref{fig3}, where for \(m = 2\) one clearly observes an additional nontrivial reordering of the fundamental mode and its first overtones.

In Fig.~\ref{fig4} we repackage these results in the complex--frequency plane by plotting the trajectories of the fundamental ($n=0$) and first two overtones ($n=1,2$) in the $(\omega_R,-\omega_I)$ portrait for $m = 1$ and $m = 2$ as the regularization scale $\ell$ is varied. At $\ell = 0$ the modes start from the exact BTZ frequencies $\omega = \pm m - 2 i M^{1/2}(n+1)$~\cite{Cardoso:2001hn}, therefore the endpoints lie at $(\omega_R,-\omega_I) = (m,2(n+1))$. As $\ell$ is increased, the corresponding curves bend monotonically towards smaller $\omega_R$ and larger $-\omega_I$, illustrating that the oscillation frequency decreases while the damping rate grows. Each trajectory terminates at a critical value $\ell = \ell_{\text{crit}}$, whose precise numerical value depends on $m$ and on the mode. The deformation is more pronounced for $m = 2$ than for $m = 1$, in agreement with Figs.~\ref{fig2} and \ref{fig3}, and confirms that higher harmonics are more sensitive to the presence of the regular core.

The bifurcation of the scalar QNM spectrum that we find for regular BTZ black holes is qualitatively similar to what has been observed for Maxwell perturbations of Schwarzschild–AdS black holes and black branes. In those systems, the Maxwell spectrum contains a branch of oscillatory modes and a branch of purely imaginary, diffusive-like modes; as the dimensionless parameter $r_+/L$ (or the wavenumber) is varied, these branches can merge and split, leading to a characteristic bifurcation and a nontrivial reordering of overtones~\cite{Miranda:2008vb,Morgan:2013dv,Wang:2021upj,Lei:2021kqv,Fortuna:2022fdd,Daghigh:2022uws}. In three dimensions the $2$-form Maxwell field is Hodge dual to a $1$-form gradient of a scalar, so it is natural that the scalar QNMs of our regular BTZ black hole share many qualitative features with Maxwell QNMs in higher-dimensional AdS backgrounds.

A closely related phenomenon occurs for scalar perturbations in Coulomb–like AdS black holes in $2+1$ dimensions, where increasing the effective charge drives the QNMs from an oscillatory regime into a purely damped regime and generates two distinct purely imaginary branches~\cite{Aragon:2021ogo}. Our regular BTZ background exhibits the same qualitative pattern: for $m>0$ and sufficiently small regularization parameter $\ell$ the scalar QNMs are oscillatory and lie on a BTZ–like complex branch (whereas for $m=0$ they already form a purely imaginary family that splits into two branches as soon as $\ell$ is switched on), while beyond a critical value $\ell_{\text{crit}}(m)$ the complex branch for $m>0$ collides with the imaginary axis and bifurcates into purely imaginary modes. In general this produces two purely imaginary branches, except for $m=2$ where a short $\ell$–interval with a residual nonzero real part induces a secondary bifurcation and yields three purely imaginary branches; in all cases, for $\ell\gtrsim \ell_{\text{crit}}(m)$ the least damped mode is purely damped.\footnote{A similar coexistence of purely imaginary and complex QNMs, often referred to as de Sitter modes and Schwarzschild/RN modes, has been observed in other contexts~\cite{Gonzalez:2022upu,Konoplya:2022xid}. Purely imaginary modes are, in fact, of particular interest in studies of QNMs and the Strong Cosmic Censorship conjecture~\cite{Cardoso:2017soq,Dias:2018ynt}.} This strongly suggests that the underlying mechanism is geometric and governed by the near-horizon structure of the effective potential, in particular by the sign and magnitude of $V''_{\text{eff}}(r_\text{h})$~\footnote{Here prime denotes derivative with respect to $r$.}, as emphasized in the Coulomb–like AdS case~\cite{Aragon:2021ogo} (see also~\cite{Myung:2008pr}). In particular, Refs.~\cite{Aragon:2021ogo, Aragon:2020qdc} showed that the transition between complex and purely imaginary quasinormal branches is controlled by a change in the concavity of the effective potential at the event horizon. For the regular BTZ black hole, this behavior is displayed in Fig.~\ref{fig5}~(left panel), where one can clearly see that the concavity of the effective potential near the horizon changes as the parameter $\ell$ is varied. For the effective potential~\eqref{veff}, the condition $V''_{\text{eff}}(r_\text{h}) = 0$ singles out the bifurcation point of the fundamental mode, which occurs at $\ell_{\text{crit}}(m) = \nicefrac{\sqrt{3}\,m}{2\sqrt{m^{2}+3}}$. The corresponding bifurcation curve in the $(\ell, m)$ plane is displayed in Fig.~\ref{fig5}~(right panel), where the black line traces the locus $V''_{\text{eff}}(r_\text{h}) = 0$ separating regions of opposite concavity. In the eikonal limit $m\to\infty$ one finds $\ell_{\rm crit}\to \sqrt{3}/2$. Thus, the bifurcation threshold saturates at large angular momentum rather than disappearing, supporting the interpretation that the branch splitting is a geometric effect tied to the deformed near-horizon potential.

At a more conceptual level, our results fit into the broader picture in which black hole QNM spectra can undergo structural transitions as one moves in parameter space. In nearly extremal Kerr spacetime, the DMs and ZDMs form two branches that bifurcate at high spin: as extremality is approached, the least damped mode switches from a DM to a ZDM branch with $\omega_I \to 0$, leading to very long-lived ringdown and power-law intermediate tails~\cite{Yang:2012pj,Yang:2013uba,Zimmerman:2015rua}. In our regular BTZ case there is no extremal limit associated with the bifurcation, and the purely imaginary branch retains a finite damping rate; nonetheless, we also observe a change in the identity of the least damped mode and a sharp reorganization of the spectrum at $\ell\simeq\ell_{\text{crit}}(m)$. This type of “mode switching’’ has been linked, in simple one-dimensional models and in Maxwell–AdS examples, to strong spectral sensitivity (or spectral instability) and to subtle features in the time-domain response, where the naive fundamental mode does not always control the dominant decay channel at late times~\cite{Li:2024npg,Gonzalez:2022upu,Ficek:2023imn}.


\section{Conclusion}\label{sec7}

We have studied the quasinormal spectrum of massless scalar fields propagating on a family of regular BTZ black holes constructed from an infinite tower of dimensionally regularized Lovelock corrections. These geometries are asymptotically AdS, reduce to the standard BTZ solution in the limit $\ell \to 0$, and resolve the central singularity by introducing a smooth core of size set by a new length scale $\ell$. Our main goal was to understand how this regularization scale imprints itself on the scalar QNMs and, in particular, whether it can induce qualitative changes in the structure of the spectrum.

The scalar QNMs were computed using both Leaver's continued–fraction method and the Horowitz--Hubeny power–series method. In the regular BTZ background, the radial equation leads to an eight–term recurrence relation for the Frobenius coefficients, which we systematically reduce to a three–term form by Gaussian elimination, thereby allowing the standard continued–fraction machinery to be applied. In the BTZ limit our numerical results reproduce the known analytic spectrum, and throughout the regular sector the two numerical schemes agree within numerical accuracy, providing a strong consistency check on the calculations.

For the regular BTZ geometries, we find that the spectrum is everywhere linearly stable: the imaginary part of the frequency satisfies $\omega_I<0$ for all parameter values explored. Quantitatively, the regularization generally reduces the magnitude $|\omega_I|$ compared to the BTZ case, thereby lengthening the ringdown time. More strikingly, as the regularization scale $\ell$ is increased, the spectrum undergoes a sequence of bifurcations. For $m=0$ the purely imaginary BTZ mode splits into two distinct purely imaginary branches. For $m\geq 1$ the BTZ–like complex branch drifts towards the imaginary axis and, at a critical value $\ell_\text{crit}(m)$, hits $\omega_R=0$ and bifurcates into multiple purely imaginary branches; for $m=2$ we find a second collision and the emergence of a third branch. In this way the identity of the least–damped mode changes as a function of $\ell$ and $m$, and the overtones are nontrivially reordered. The resulting pattern closely resembles the quasinormal–mode bifurcations known from nearly extremal Kerr black holes and from Maxwell and scalar perturbations of asymptotically AdS black holes and black branes, indicating that regular BTZ black holes provide a simple three–dimensional laboratory for such phenomena.

Finally, it is worth noting that the regularization does not modify the leading near-horizon scalar dynamics. Writing $r=r_{\mathrm h}+x$ and taking the limit $x\to 0$, the radial Klein--Gordon equation reduces at leading order to
\[
R''(x)+\frac{1}{x}R'(x)+\frac{L^{4}\omega^{2}}{4r_{\mathrm h}^{2}}\frac{1}{x^{2}}\,R(x)=0 .
\]
Thus, the leading near-horizon equation is identical to that of the BTZ black hole. After the field redefinition $\chi=\sqrt{x}\,R$, the equation takes the universal inverse-square form
\[
H\chi=\left[-\frac{d^{2}}{dx^{2}}+\frac{a}{x^{2}}\right]\chi,
\qquad
a=-\left(\frac{1}{4}+\frac{\omega^{2}}{A^{2}}\right),
\]
with the near-horizon expansion
\[
f(x)\sim A x\,\bigl[1+\mathcal{O}(x)\bigr], \qquad A=\frac{2r_{\mathrm h}}{L^{2}} .
\]
This is precisely the structure discussed in Ref.~\cite{Chakrabarti:2007sj} for a wide class of black holes, and in the present case the coefficient $A$ is again the same as in the BTZ geometry. The $\ell$-dependent corrections enter only at subleading order. Therefore, although the regularization substantially modifies the global quasinormal spectrum and produces the bifurcation pattern discussed above, it preserves the universal leading near-horizon conformal structure of the BTZ geometry.

It is also worth noting that the regularization does not alter the Hawking temperature at the outer horizon. Since the event horizon remains at $r_{\mathrm h}=L\sqrt{M}$, the temperature is still
\[
T_H=\frac{f'(r_{\mathrm h})}{4\pi}=\frac{r_{\mathrm h}}{2\pi L^2},
\]
exactly as in the BTZ case~\cite{Fernandes:2025eoc}. This is consistent with our near-horizon analysis, where the leading coefficient in the expansion $f(x)\sim A x$ is unchanged, $A=2r_{\mathrm h}/L^2$, so that the scalar field probes the same leading near-horizon conformal structure as in BTZ. In this sense, unlike many regular black-hole models where one encounters a zero-temperature extremal outer horizon~\cite{Kovacik:2022ngv}, the present regularization leaves the outer-horizon thermodynamics unchanged; the extremal feature instead appears at the inner horizon, which has vanishing surface gravity.

It is also interesting to view our results in the broader context of earlier BTZ-like studies, where the emergence of branch splitting or oscillatory-to-purely-imaginary transitions has been reported~\cite{Aragon:2021ogo, DalBoscoFontana:2023syy, Gonzalez:2021vwp, deOliveira:2024pab}. Here we have provided a dedicated and more detailed characterization of the phenomenon by tracking multiple overtones, identifying repeated bifurcations and branch collisions, and clarifying how mode ordering and the least-damped sector can change across parameter space. A definitive statement about generality or universality, however, calls for comparably detailed surveys in other regular BTZ models (for instance those arising from nonlinear electrodynamics or other higher-curvature completions), performed with the same level of spectral resolution. Finally, it would be valuable to explore possible dimensional dependence by extending these analyses beyond three dimensions, in particular to four-dimensional regular black holes where the perturbation structure and mode families can be richer.

In this paper we have focused exclusively on scalar QNMs. A natural extension is to analyze the perturbations of other fields (such as Dirac or Maxwell perturbations) on the same regular BTZ background. We have also restricted ourselves to the simplest regular BTZ configuration, corresponding to a particular choice of the Lovelock coefficients $c_n$; it would be interesting to investigate how the quasinormal spectrum and its bifurcation pattern change for other regular geometries obtained by varying the $c_n$. Although Leaver's continued–fraction method is particularly accurate and robust, a complementary, more systematic study of overtones could be carried out using alternative techniques such as pseudospectral methods. Although Leaver's continued–fraction method is particularly accurate and robust for individual overtones, a complementary and more systematic mapping of the overtone structure would be naturally carried out using Chebyshev-based pseudospectral techniques. The main practical advantage of such methods in the present context is that a single diagonalization of the discretized eigenvalue problem returns the fundamental mode together with many overtones simultaneously, without the need for individual root searches and initial guesses, a feature that is particularly useful when scanning a continuous parameter, such as the regularization scale $\ell$, across regions where modes bifurcate, collide and reorder. A further advantage, especially relevant for the extension to massive scalar and Dirac perturbations mentioned above, is that working in ingoing Eddington–Finkelstein or hyperboloidal coordinates reduces the QNM boundary conditions to regularity requirements that are automatically enforced on the collocation grid: the leading near-horizon and near-boundary exponents and the explicit recurrence coefficients, which in a Leaver-type or Horowitz–Hubeny treatment must be rederived for arbitrary mass, no longer need to be obtained analytically beforehand. The same framework also provides the natural setting for pseudospectrum and spectral-(in)stability analyses of regular BTZ black holes, along the lines pursued for asymptotically flat and AdS spacetimes in~\cite{Jaramillo:2020tuu, Destounis:2021lum, Arean:2023ejh, Konoplya:2023kem}. We leave a detailed pseudospectral and pseudospectrum-based study of the bifurcation phenomenon reported here for future work. Finally, the Leaver–type implementation developed here can be readily adapted to other $(2+1)$–dimensional black hole spacetimes.

\section*{Acknowledgement}
This research was supported by the Croatian Science Foundation Project No. IP-2025-02-8625, \emph{Quantum aspects of gravity}.

\appendix

\section{Coefficients of recurrence relation and Gauss elimination procedure}\label{app1}
In this appendix we list the coefficients of recurrence relation and detail the Gauss elimination procedure used to reduce the eight-term recurrence relation to a three-term recurrence relation. The explicit expression for the coefficients appearing in the equation \eqref{sevenrecurrence} are,
\begin{widetext}
\begin{equation}
\begin{aligned}
    \alpha _n = & \rho ^2 - 4 r_\text{h}^2 (b \rho + n)^2 \\
    \beta _n = & 2 r_\text{h}^2 \left(4 b^2 \rho^2 + b \rho (8 n - 1) + n (4 n - 1) + 1\right) + \ell^2 \left(8 r_\text{h}^2 (b \rho + n - 2)(b \rho + n - 1) - 6 \rho ^2\right) + 2 m^2 - \rho ^2 \\
    \gamma _n = & \ell^2 \left(-4 r_\text{h}^2 (b \rho + n - 2)(5 b \rho + 5 n - 9) - 8 m^2 + 9 \rho ^2\right) - r_\text{h}^2 \left(5 \left(2 b \rho n + (b \rho - 1) b \rho + n^2\right) - 5 n + 2\right) - 3 m^2 + 12 \ell^4 \rho ^2 \\
    \delta _n = & \ell^2 \left(18 r_\text{h}^2 (b \rho + n - 3)(b \rho + n - 2) + 16 m^2 - 3 \rho ^2\right) + r_\text{h}^2 (b \rho + n - 1)^2 + m^2 + 8 \ell^4 (m^2 - 3 \rho ^2) - 8 \ell^6 \rho ^2 \\
    \zeta _n = & \ell^2 \left(-r_\text{h}^2 (b \rho + n - 4)(7 b \rho + 7 n - 17) - 10 m^2\right) + \ell^4 (15 \rho ^2 - 20 m^2) + 20 \ell^6 \rho ^2 \\
    \eta _n = & \ell^2 \left(r_\text{h}^2 (b \rho + n - 5)(b \rho + n - 3) + 2 m^2\right) + \ell^4 (18 m^2 - 3 \rho ^2) - 18 \ell^6 \rho ^2 \\
    \theta_n = & 7 \ell^6 \rho ^2 - 7 m^2 \ell^4 \\
    \tau _n = & m^2 \ell^4 - \ell^6 \rho ^2.
\end{aligned}
\end{equation}
\end{widetext}

\begin{table*}[ht!]
\centering
\vspace{0.5cm}
\caption{Analytic and numerical (Leaver) QNM frequencies for BTZ black holes.  Entries show ``Analytic / Numerical''. }
\label{tab2}

\begin{tabular}{c|c|c|c|c|c}
\hline\hline
$m \backslash n$ 
& $0$ & $1$ & $2$ & $5$ & $10$ \\
\hline\hline

0 
& $0 - 2i$ / $0.000 - 2.000i$
& $0 - 4i$ / $0.000 - 4.000i$
& $0 - 6i$ / $0.000 - 6.000i$
& $0 - 12i$ / $0.000 - 12.000i$
& $0 - 22i$ / $0.000 - 22.000i$
\\[6pt]
\hline

1 
& $1 - 2i$ / $1.000 - 2.000i$
& $1 - 4i$ / $1.000 - 4.000i$
& $1 - 6i$ / $1.000 - 6.000i$
& $1 - 12i$ / $1.000 - 12.000i$
& $1 - 22i$ / $1.000 - 22.000i$
\\[6pt]
\hline

2 
& $2 - 2i$ / $2.000 - 2.000i$
& $2 - 4i$ / $2.000 - 4.000i$
& $2 - 6i$ / $2.000 - 6.000i$
& $2 - 12i$ / $2.000 - 12.000i$
& $2 - 22i$ / $2.000 - 22.000i$
\\[6pt]

\hline\hline
\end{tabular}

\vspace{0.5cm}
\end{table*}

Now we detail the Gaussian-elimination procedure used to reduce the eight-term recurrence relation \eqref{sevenrecurrence} to a three-term recurrence. Each
elimination sweep removes the farthest-back term and reduces the order by one; therefore, starting from an eight-term recurrence, exactly five
sweeps are necessary and sufficient to reach the three-term recurrence employed in the continued-fraction condition. 

\noindent\emph{Sweep 1: eliminate $\tau_n$ (for $n\ge 7$).}
For $n=1,\ldots,6$ the $\tau_n a_{n-7}$ term is absent, hence we set
\begin{eqnarray}
\alpha_n'=\alpha_n,\ \beta_n'=\beta_n,\ \gamma_n'=\gamma_n,\ \delta_n'=\delta_n, \nonumber\\
\zeta_n'=\zeta_n,\ \eta_n'=\eta_n,\ \theta_n'=\theta_n.
\nonumber
\end{eqnarray}
For $n\ge 7$ we eliminate $\tau_n$ recursively via
\begin{eqnarray}
\alpha_n'=\alpha_n,\qquad
\beta_n'=\beta_n-\frac{\alpha_{n-1}'\,\tau_n}{\theta_{n-1}'}, \nonumber\\
\gamma_n'=\gamma_n-\frac{\beta_{n-1}'\,\tau_n}{\theta_{n-1}'},\qquad
\delta_n'=\delta_n-\frac{\gamma_{n-1}'\,\tau_n}{\theta_{n-1}'}, \nonumber\\
\zeta_n'=\zeta_n-\frac{\delta_{n-1}'\,\tau_n}{\theta_{n-1}'},\qquad
\eta_n'=\eta_n-\frac{\zeta_{n-1}'\,\tau_n}{\theta_{n-1}'}, \nonumber\\
\theta_n'=\theta_n-\frac{\eta_{n-1}'\,\tau_n}{\theta_{n-1}'}.\nonumber
\end{eqnarray}
After this sweep, the recurrence can be written as
\begin{equation}
\begin{aligned}
\alpha_n' a_n +\beta_n' a_{n-1}+ \gamma_n' a_{n-2}  +  \delta_n' a_{n-3}  + \zeta_n' a_{n-4} \\ +\eta_n' a_{n-5}+\theta_n' a_{n-6} =0.
\end{aligned}
\end{equation}

\noindent\emph{Sweep 2: eliminate $\theta_n'$ (for $n\ge 6$).}
For $n=1,\ldots,5$ the $\theta_n' a_{n-6}$ term is absent, so
\begin{eqnarray}
\alpha_n''=\alpha_n',\ \beta_n''=\beta_n',\ \gamma_n''=\gamma_n', \nonumber\\
\delta_n''=\delta_n',\ \zeta_n''=\zeta_n',\ \eta_n''=\eta_n'.\nonumber
\end{eqnarray}
For $n\ge 6$ we eliminate $\theta_n'$ via
\begin{eqnarray}
\alpha_n''=\alpha_n',\qquad
\beta_n''=\beta_n'-\frac{\alpha_{n-1}''\,\theta_n'}{\eta_{n-1}''},\nonumber\\
\gamma_n''=\gamma_n'-\frac{\beta_{n-1}''\,\theta_n'}{\eta_{n-1}''},\qquad
\delta_n''=\delta_n'-\frac{\gamma_{n-1}''\,\theta_n'}{\eta_{n-1}''},\nonumber\\
\zeta_n''=\zeta_n'-\frac{\delta_{n-1}''\,\theta_n'}{\eta_{n-1}''},\qquad
\eta_n''=\eta_n'-\frac{\zeta_{n-1}''\,\theta_n'}{\eta_{n-1}''}.\nonumber
\end{eqnarray}
This yields the six-term recurrence
\begin{equation}
\begin{aligned}
\alpha_n''a_n+\beta_n''a_{n-1}+\gamma_n''a_{n-2}+\delta_n''a_{n-3}\\
+\zeta_n''a_{n-4}+\eta_n''a_{n-5}=0.
\end{aligned}
\end{equation}

\noindent\emph{Sweep 3: eliminate $\eta_n''$ (for $n\ge 5$).}
For $n=1,\ldots,4$ set
\begin{eqnarray}
\alpha_n^{(3)}=\alpha_n'',\ \beta_n^{(3)}=\beta_n'',\ \gamma_n^{(3)}=\gamma_n'', \nonumber\\
\delta_n^{(3)}=\delta_n'',\ \zeta_n^{(3)}=\zeta_n''.\nonumber
\end{eqnarray}
For $n\ge 5$ eliminate $\eta_n''$ via
\begin{eqnarray}
\alpha_n^{(3)}=\alpha_n'',\qquad
\beta_n^{(3)}=\beta_n''-\frac{\alpha_{n-1}^{(3)}\,\eta_n''}{\zeta_{n-1}^{(3)}}, \nonumber\\
\gamma_n^{(3)}=\gamma_n''-\frac{\beta_{n-1}^{(3)}\,\eta_n''}{\zeta_{n-1}^{(3)}},\qquad
\delta_n^{(3)}=\delta_n''-\frac{\gamma_{n-1}^{(3)}\,\eta_n''}{\zeta_{n-1}^{(3)}}, \nonumber\\
\zeta_n^{(3)}=\zeta_n''-\frac{\delta_{n-1}^{(3)}\,\eta_n''}{\zeta_{n-1}^{(3)}}. \nonumber
\end{eqnarray}
This yields the five-term recurrence
\begin{equation}
\begin{aligned}
\alpha_n^{(3)}a_n+\beta_n^{(3)}a_{n-1}+\gamma_n^{(3)}a_{n-2}
+\delta_n^{(3)}a_{n-3} \\ +\zeta_n^{(3)}a_{n-4}=0.
\end{aligned}
\end{equation}

\noindent\emph{Sweep 4: eliminate $\zeta_n^{(3)}$ (for $n\ge 4$).}
For $n=1,\ldots,3$ set
\begin{eqnarray}
\alpha_n^{(4)}=\alpha_n^{(3)},\ \beta_n^{(4)}=\beta_n^{(3)},\ \gamma_n^{(4)}=\gamma_n^{(3)} ,\ \delta_n^{(4)}=\delta_n^{(3)} .\nonumber
\end{eqnarray}
For $n\ge 4$ eliminate $\zeta_n^{(3)}$ via
\begin{eqnarray}
\alpha_n^{(4)}=\alpha_n^{(3)},\qquad
\beta_n^{(4)}=\beta_n^{(3)}-\frac{\alpha_{n-1}^{(4)}\,\zeta_n^{(3)}}{\delta_{n-1}^{(4)}}, \nonumber\\
\gamma_n^{(4)}=\gamma_n^{(3)}-\frac{\beta_{n-1}^{(4)}\,\zeta_n^{(3)}}{\delta_{n-1}^{(4)}},\qquad
\delta_n^{(4)}=\delta_n^{(3)}-\frac{\gamma_{n-1}^{(4)}\,\zeta_n^{(3)}}{\delta_{n-1}^{(4)}} .\nonumber
\end{eqnarray}
This yields the four-term recurrence
\begin{equation}
\alpha_n^{(4)}a_n+\beta_n^{(4)}a_{n-1}+\gamma_n^{(4)}a_{n-2}
+\delta_n^{(4)}a_{n-3}=0.
\end{equation}

\noindent\emph{Sweep 5: eliminate $\delta_n^{(4)}$ (for $n\ge 3$).}
For $n=1,2$ set
\begin{equation}
\alpha_n^{(5)}=\alpha_n^{(4)},\qquad \beta_n^{(5)}=\beta_n^{(4)},\qquad \gamma_n^{(5)}=\gamma_n^{(4)}.
\end{equation}
For $n\ge 3$ eliminate $\delta_n^{(4)}$ via
\begin{eqnarray}
\alpha_n^{(5)}=\alpha_n^{(4)},\qquad
\beta_n^{(5)}=\beta_n^{(4)}-\frac{\alpha_{n-1}^{(5)}\,\delta_n^{(4)}}{\gamma_{n-1}^{(5)}},\nonumber\\
\gamma_n^{(5)}=\gamma_n^{(4)}-\frac{\beta_{n-1}^{(5)}\,\delta_n^{(4)}}{\gamma_{n-1}^{(5)}}.\nonumber
\end{eqnarray}
This produces the desired three-term recurrence
\begin{equation}
\alpha_n^{(5)}a_n+\beta_n^{(5)}a_{n-1}+\gamma_n^{(5)}a_{n-2}=0,
\label{eq-threetermapp}
\end{equation}
which is the form used in the continued-fraction condition.

\section{Leaver's method for BTZ black hole} \label{app2}
In this appendix we detail the application of Leaver's method to BTZ black hole. We compare our results with the QNM obtained from the analytical expression \cite{Cardoso:2001hn}. The asymptotic behavior of the \ref{Waveradial} for the BTZ potential has the same form as given in \ref{scalarbc} with 
\begin{equation}
b=\frac{L^2}{2r_\text{h}}.
\label{bvaluebtz}
\end{equation}
Using the ansatz \ref{ansatz} in \ref{Waveradial} for the scalar perturbation in the BTZ black hole spacetime we obtain a simpler four-term recurrence relation with the recurrence coefficients, 

\begin{equation}
\begin{aligned}
    \alpha _n = & \rho^2 - 4 r_\text{h}^2 (b \rho + n)^2 \\
    \beta _n = & 2 r_\text{h}^2 \left(4 b^2 \rho^2 + b \rho (8 n - 1) + n (4 n - 1) + 1\right) + 2 m^2 - \rho^2 \\
    \gamma _n = & r_\text{h}^2 \left(-\left(5 \left(2 b \rho n + (b \rho - 1) b \rho + n^2\right) - 5 n + 2\right)\right) - 3 m^2 \\
    \delta _n = & r_\text{h}^2 (b \rho + n - 1)^2 + m^2
\end{aligned}
\end{equation}

Applying Gauss elimination procedure once we obtain a three-term recurrence relation and QNM frequencies are calculated as described in section \ref{sec4}. The values are in good agreement with the analytical results~\cite{Cardoso:2001hn} given by the expression $\omega = \pm m -2 i M ^{1/2} (n+1)$, which are tabulated in table~\ref{tab2}.

\bibliography{reference}
\end{document}